\documentclass[aps,prl,amsmath,amssymb,floatfix,twocolumn, amsmath, superscriptaddress, twocolumn]{revtex4}
\usepackage{multirow}
\usepackage{bbold}
\usepackage{subfigure}
\usepackage{color}
\usepackage{mathrsfs}
\usepackage{hyperref}

\usepackage{amsfonts, relsize, color}
\usepackage{graphics}
\usepackage{graphicx}
\usepackage{subfigure}
\usepackage{hyperref}
\usepackage{color}

\begin{document}
\title{Collisionless Transport Close to a Fermionic Quantum Critical Point in Dirac Materials}

\author{Bitan Roy}
\affiliation{Max-Planck-Institut f\"{u}r Physik komplexer Systeme, N\"{o}thnitzer Stra. 38, 01187 Dresden, Germany}

\author{Vladimir Juri\v ci\' c}
\affiliation{Nordita, KTH Royal Institute of Technology and Stockholm University, Roslagstullsbacken 23,  10691 Stockholm,  Sweden}

\date{\today}

\begin{abstract}
Quantum transport close to a critical point is a fundamental, but enigmatic problem due to fluctuations, persisting at all length scales. We report the scaling of optical conductivity (OC) in the \emph{collisionless} regime ($\hbar \omega \gg k_B T$) in the vicinity of a relativistic quantum critical point, separating two-dimensional ($d=2$) massless Dirac fermions from a fully gapped insulator  or superconductor. Close to such critical point gapless fermionic and bosonic excitations are strongly coupled, leading to a \emph{universal} suppression of the inter-band OC as well as of the Drude peak (while maintaining its delta function profile) inside the critical regime, which we compute to the leading order in $1/N_f$- and $\epsilon$-expansions, where $N_f$ counts fermion flavor number and $\epsilon=3-d$. Correction to the OC at such a non-Gaussian critical point due to the long-range Coulomb interaction and generalizations of these scenarios to a strongly interacting  three-dimensional Dirac or Weyl liquid are also presented, which can be tested numerically  and possibly from 
non-pertubative gauge-gravity duality, for example. 
\end{abstract}

\maketitle

\emph{Introduction}. Quantum critical phenomena in strongly interacting low-dimensional itinerant fermionic systems have attracted ample attention in recent time~\cite{sachdev-book}. Such a broad arena can be divided into two sectors, namely when long-lived gapless fermionic excitations reside around (a) a few isolated points in the Brillouin zone (referred as Fermi points), or (b) a closed contour in the reciprocal space, the Fermi surface, with our focus being solely on the former system. A paradigmatic representative of a \emph{nodal Fermi liquid} is constituted by quasi-relativistic Dirac excitations, which find its condensed matter realization, for example, in graphene~\cite{graphene-RMP} and on the surface of topological insulators~\cite{TI-review-1, TI-review-2}.

\begin{figure}[t!]
\includegraphics[width=5cm,height=5cm]{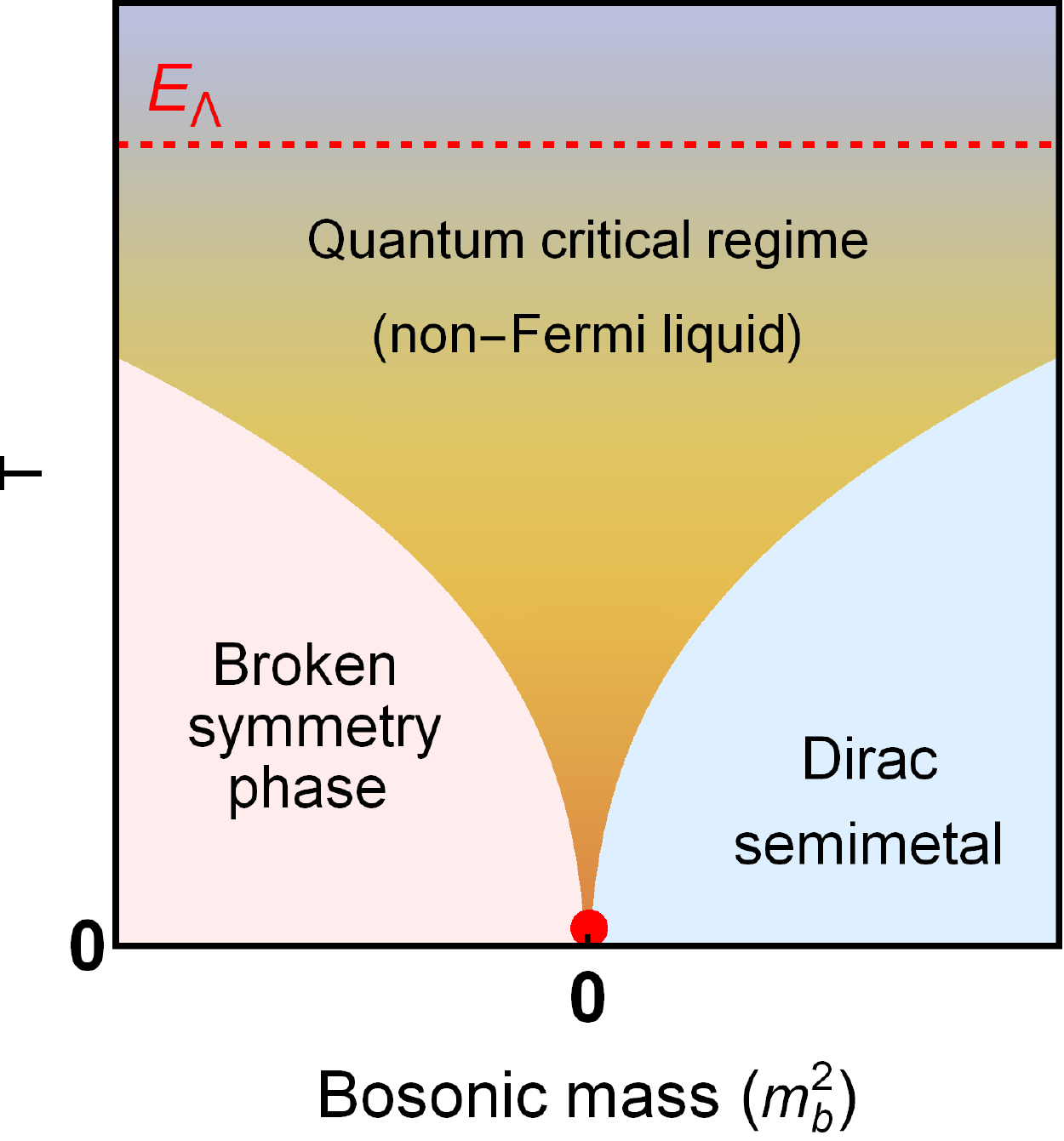}
\caption{ A schematic representation of the quantum phase transition from a Dirac semimetal to a gapped ordered phase through Gross-Neveu-Yukawa QCP (red dot) located at $m^2_b=0$ and the associated quantum critical fan (shaded region) at finite temperature. Here, $m^2_b$ is the bosonic mass and the critical regime is occupied by a non-Fermi liquid. The scaling of OC in this regime is given by Eq.~(\ref{interband-final}) [inter-band component] and Eq.~(\ref{Drude-final}) [Drude component]. The quantum critical scaling ceases to operate at a (non-universal) high energy scale $E_\Lambda \sim 1$eV in a graphene-like system (red dotted line), yielding a wide window of frequency ($\omega<2E_\Lambda$) over which our proposed universal scaling of OC remains operative.
}~\label{criticalfan}
\end{figure}

A two dimensional Dirac system can undergo continuous quantum phase transitions into a plethora of Mott insulators, such as antiferromagnet~\cite{herbut-original, herbut-juricic-vafek, sorella-1, assaad-herbut-1, assaad-herbut-2, sorella-2} and charge-density wave~\cite{herbut-original,HJR-PRB, troyer, hong-yao-1} (both relevant for graphene), or superconducting phases, for example, the $s$-wave pairing (relevant for graphene and surface states of a topological insulator)~\cite{honerkamp, paramekanti, roy-herbut-kekule}, at strong coupling depending on the relative strength of short-range repulsive or attractive interactions. By now there exists compelling evidence that the associated quantum critical behavior can be captured by an effective Gross-Neveu-Yukawa (GNY) field theory that besides standard order-parameter fluctuations, also accounts for the coupling between gapless fermionic excitations and bosonic order-parameter field~\cite{zinn-justin, rosenstein, herbut-juricic-vafek, roy-MCP, roy-juricic-herbut, roy-juricic-MCP, classen-herbut-scherrer, herbut-MCP, RGJ-multicriticality, sorella-1, assaad-herbut-1, assaad-herbut-2, sorella-2, troyer, hong-yao-1,hong-yao-2, kaul}. Concomitantly, the interacting GNY quantum critical point (QCP) and the corresponding critical regime, shown in Fig.~\ref{criticalfan}, host a strongly coupled \emph{non-Fermi liquid}, where the notion of any sharp quasi-particle excitations becomes moot.

The question arises how to theoretically understand possible experimental ramifications of such  a strongly coupled relativistic non-Fermi liquid and in this Letter we present its imprint on the optical conductivity (OC). So far much focus has been on the OC at purely bosonic QCPs, as in the case of superconductor-insulator transition~\cite{fisher, cha, wallin, fazio, herbut, sorensen, krempa, prokofiev, aurbach}, at a very specific supersymmetric QCP~\cite{krempa-maciejko} or spin-fermion model~\cite{abanov}. We here reveal universal features of the quantum transport at the generic strongly coupled fermionic QCP in two spatial dimensions, separating a Dirac liquid and an interaction-driven gapped state. Note that in pure bosonic systems universal and finite conductivity (due to gapless bosonic excitations of charge $2e$) can only be found at the interacting QCP~\cite{fisher, cha, wallin, fazio, herbut, sorensen, krempa, prokofiev, aurbach}. In contrast, our analysis establishes a universal suppression of the OC at the GNY QCP and in the associated non-Fermi liquid, in comparison to that of the non-interacting nodal Dirac liquid (accommodating only gapless fermionic excitations of charge $e$). Its physical origin lies in a strong coupling between quantum critical fermionic and bosonic excitations, falling outside the paradigm of the standard purely bosonic $\Phi^4$ theory.

Our analysis relies on a perturbative method, controlled by the distance from the upper-critical three dimensions of the theory (an $\epsilon$-expansion) and fermionic flavor number ($1/N_f$-expansion). On the same token, we address the quantum critical transport in three dimensions, and the influence of the long-range Coulomb interaction (always present in a real system) at a fermionic critical point. Our key results can be summarized as follows: We find universal suppression of both inter-band [see Eq.~(\ref{interband-final})] and Drude [see Eq.~(\ref{Drude-final})] components of the OC near the GNY QCP in conjunction with the enhancement of the former piece by the long-range tail of Coulomb interaction [see Eqs.~(\ref{coulomb_general}) and (\ref{coulomb3D})].

The gauge invariance assures that conductivity ($\sigma$) scales as $\sigma \sim L^{d-2}$ with the system size $L$. Therefore, in two spatial dimensions ($d=2$) conductivity at finite temperature ($T$) and frequency ($\omega$) is a \emph{universal} function of the dimensionless ratio $x=\hbar \omega/(k_B T)$, namely $\sigma=\frac{e^2}{h} f(x)$, where $e^2/h$ is the quantum of conductance.
For $\hbar \omega \gg k_B T$ \emph{collisionless} transport is dominated by coherent excitations created by the external electric field.
By contrast, in the high-temperature limit ($k_B T \gg \hbar \omega$), also known as \emph{collision-dominated} or \emph{hydrodynamic} regime, a plasma of pre-existing thermal excitations, while achieving local equilibration via incoherent mutual collisions, dominates transport, and typically $f(\infty)<f(0)$~\cite{sachdev-book, damle-sachdev}.

\begin{figure}[t!]
\subfigure[]{
\includegraphics[width=4cm,height=3.5cm]{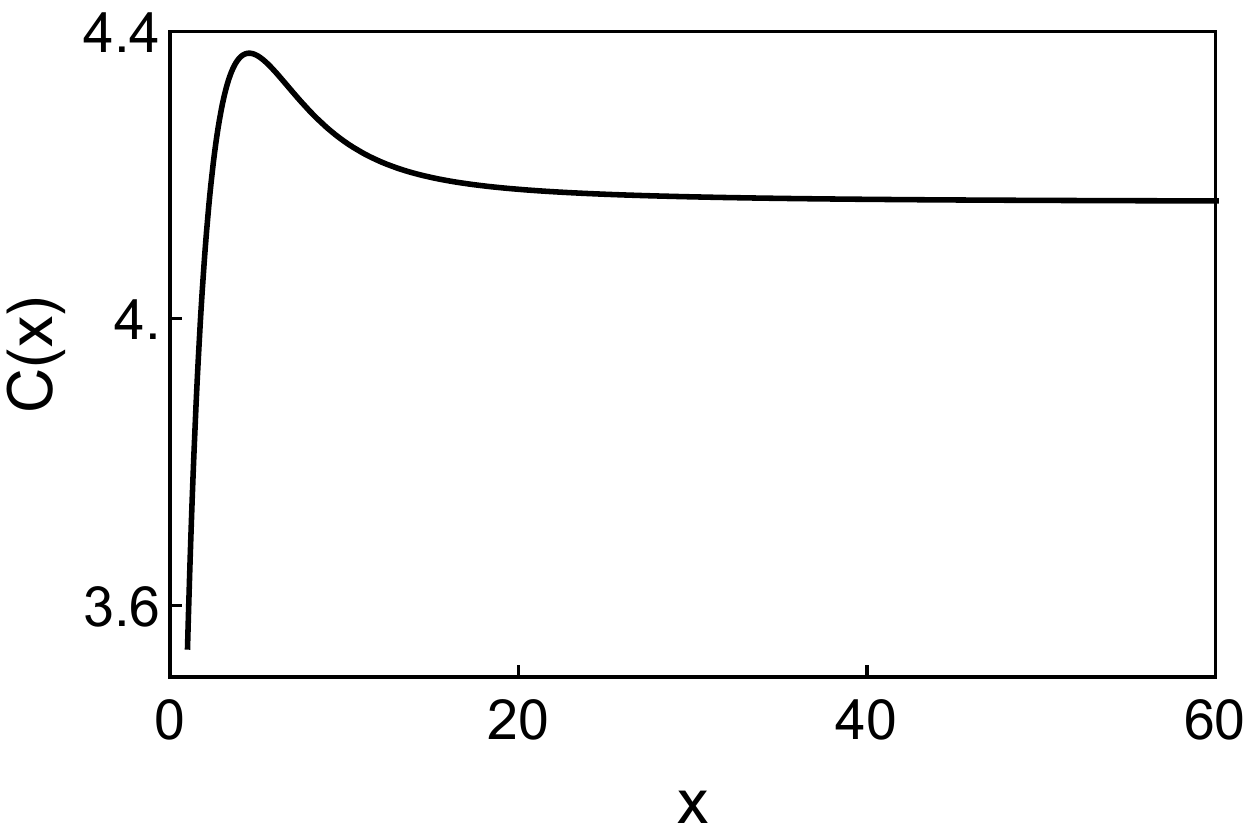}~\label{fig:funcC}
}
\subfigure[]{
\includegraphics[width=4cm,height=3.5cm]{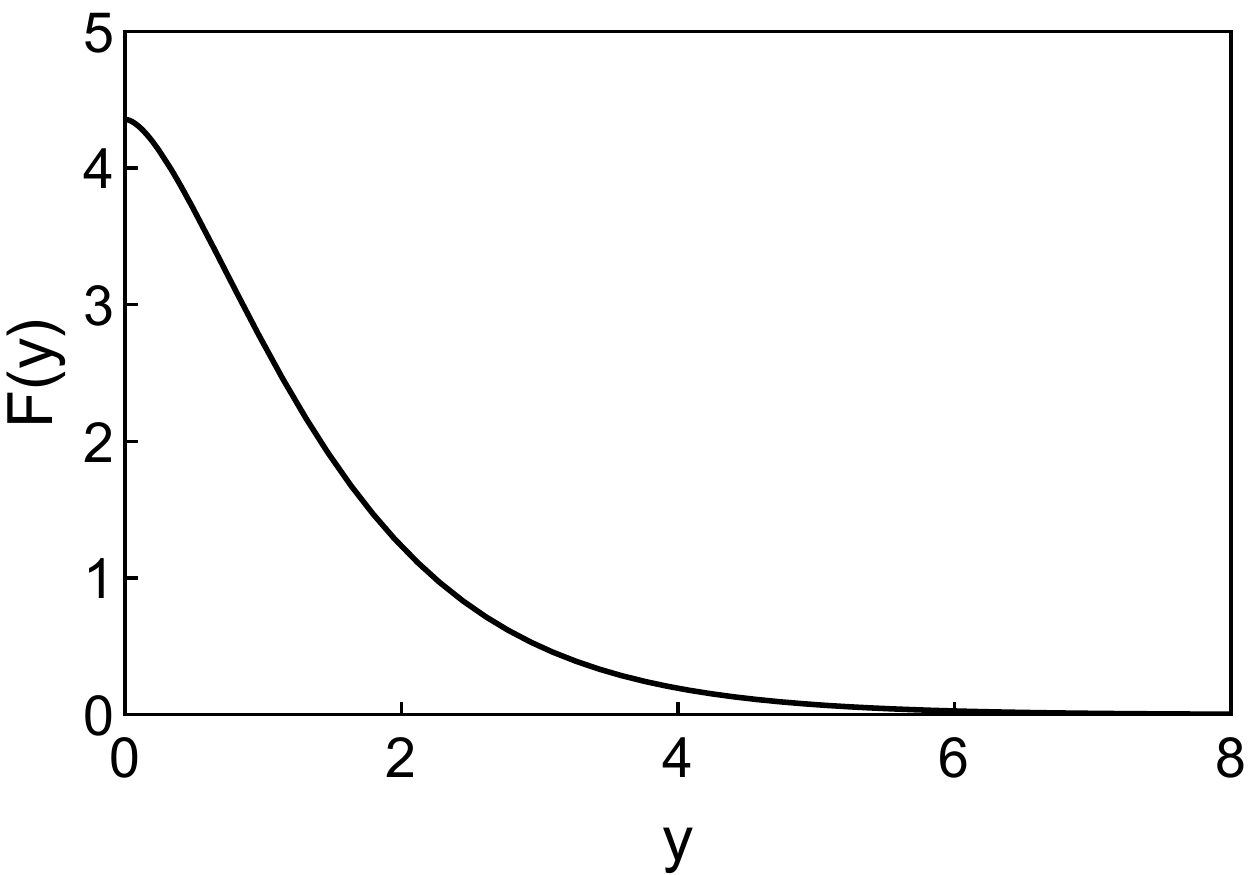}~\label{fig:funcF}
}
\caption{ Scaling of two universal functions (a) $C(x)$ and (b) $F(y)$, respectively governing the suppression of inter-band [see Eq.~(\ref{interband-final})] and Drude [see Eq.~(\ref{Drude-final})] components of the OC in the vicinity of the fermionic critical point in two dimensions. Here, $x=\hbar \omega/(k_BT)$ and $y=m_f(T)/(k_BT)$.
}
\end{figure}

We first show that the inter-band piece of the OC at the relativistic GNY QCP is given by~\cite{comment}
\begin{equation}~\label{interband-final}
\sigma^{\rm IB}_\ast \left( x \right)=  \left[ 1-\frac{N_b}{2N_f} \; \left[ 1+\epsilon \; C\left(x \right) \right] \right]
\sigma_0^{\rm IB}\left( x \right),
\end{equation}
in the collisionless regime, to the leading order in $\epsilon$- and $1/N_f$-expansions, and for physical situation $d=2$ or equivalently $\epsilon=1$. Here, $\sigma^{\rm IB}_0(x)=(N_f\pi/4)$ $\tanh(x/4)$ is the inter-band OC [in units of $e^2/h$, set to one hereafter] of a noninteracting Dirac liquid, $N_f$ is the number of four-component Dirac fermion species (hence, for graphene and surface states of toplogical insulators $N_f=2$ and $1/2$, respectively), and $N_b$ counts the number of real order-parameter components. The scaling of the universal function $C(x)$ is displayed in Fig.~\ref{fig:funcC}. Next, we show that inside the critical regime the Drude part scales as
\begin{equation}~\label{Drude-final}
\sigma^{\rm D}_\ast \left( x \right)= N_f F \left[G(N_f,N_b)\sqrt{\epsilon}\;\right] \; \delta\left( x \right),
\end{equation}
with $\delta(x)$ as the Dirac delta function, and $F$ and $G$ are two universal functions of their arguments, about which in a moment. The Drude peak for the noninteracting system [$\sigma^{\rm D}_0 \left( x \right)$] is recovered by setting $\epsilon=0$, for which $F(0)=2\pi \ln 2$ and $\sigma^{\rm D}_0 \left( x \right)=2 N_f \pi \ln 2 \; \delta\left( x \right)$. Otherwise, $F(y)$ is a positive-definite and monotonically decreasing function, see Fig.~\ref{fig:funcF}. Therefore, due to a strong interaction between the gapless fermionic and bosonic degrees of freedom, the OC inside the quantum critical regime gets reduced in comparison to its counterpart in a non-interacting Dirac fluid. On the insulating side of the transition OC displays \emph{activated} behavior.

\emph{GNY Model}. We first briefly review the quintessential features of the critical GNY theory in $d+1$ space-(imaginary) time dimensions, with the Euclidean action ${\mathcal S}=\int d\tau d^d{\bf r} \; (L_f+L_Y+L_b)$, describing massless Dirac fermions coupled with the critical bosonic excitations corresponding to an $O(N_b)$ symmetry breaking order-parameter. The fermionic Lagrangian is given by
\begin{equation}
L_f=\Psi^\dagger(\tau, {\bf r}) \; [ \partial_0-i\sum_{j=1}^{d} \; \Gamma_j \partial_j ] \; \Psi(\tau, {\bf r}).
\end{equation}
The $\Gamma$ matrices satisfy the anti-commuting Clifford algebra $\{ \Gamma_j, \Gamma_k \}=2 \delta_{jk}$. 
The coupling between fermionic and bosonic excitations is captured by
\begin{equation}
L_{Y}=g\sum^{N_b}_{\alpha=1} \Phi_\alpha(\tau, {\bf r}) \Psi^\dagger(\tau, {\bf r}) M_\alpha \Psi (\tau, {\bf r}),
\end{equation}
with $g$ as the Yukawa coupling constant. Here, $\Gamma_j$s and $M_\alpha$s are $8\times 8$ Hermitian matrices, satisfying $\left\{ \Gamma_j,M_\alpha \right\}=0$.
The $O(N_b)$ symmetric purely bosonic action reads as
\begin{equation}
L_b=\sum_{\alpha=1}^{N_b} \left[ \frac{1}{2}\Phi_\alpha \left( -\sum^{d}_{\mu=0}\partial_\mu^2+m^2_b \right) \Phi_\alpha + \frac{\lambda}{4!} [\Phi_\alpha^2]^2 \right],
\end{equation}
with $m^2_b$ as the tuning parameter for the transition, equal zero at the QCP, $\lambda$ is the four-boson coupling, and $\Phi_\alpha \equiv \Phi_\alpha(\tau, {\bf r})$. The Fermi ($v_F$) and bosonic ($v_b$) velocities are assumed to be the same, due to the \emph{emergent} Lorentz symmetry, and set to be \emph{unity} throughout~\cite{RJH-lorentz}.

Since both Yukawa ($g$) and the four boson ($\lambda$) couplings are marginal in $d=3$, the flow of these two couplings can be controlled by an $\epsilon$-expansion about three spatial dimensions. To the leading order in $\epsilon$, the renormalization group flow equations are given by
\begin{eqnarray}~\label{RG:zippedgeneral}
\beta_{g^2} &=& \epsilon g^2- (2N_f+4-N_b) g^4, \nonumber \\
\beta_{\lambda} &=& \epsilon \lambda -4 N_f g^2 \left(\lambda-6 g^2 \right) - \frac{\lambda^2}{6} \left( 8+N_b\right),
\end{eqnarray}
in the critical hyperpplane defined by $m^2_b=0$, in terms of dimensionless coupling constants $X q^{-\epsilon}/(8 \pi^2)$ $\to X$ for $X=g^2, \lambda$. Here $q$ is a momentum scale defining the \emph{infrared} renormalization-group $\beta-$function for a coupling $X$ as $\beta_X\equiv-dX/d\ln q$. The above coupled flow equations support \emph{only one fully stable} fixed point located at
\begin{equation}~\label{FP:locationgeneral}
\left( g^2_\ast, \lambda_\ast \right)= \left( 1, \frac{3}{a_3} \left[ a_2+ \sqrt{a_2^2+ 16 N_f a_3}\right]\right) \frac{\epsilon}{a_1},
\end{equation}
also known as GNY critical point, where $a_1=2 N_f+4-N_b$, $a_2=4-2N_f-N_b$ and $a_3=N_b+8$. At this QCP both fermionic and bosonic excitations possess non-trivial anomalous dimensions, respectively given by $\eta_f=N_b g^2_\ast/2$ and $\eta_b= 2 N_f g^2_\ast$, responsible for the absence of sharp quasiparticles in its vicinity. The associated quantum critical fan thus accommodates a \emph{non-Fermi liquid}. Also the ratio of the fermionic ($m^2_f$) and bosonic ($m^2_b$) masses assumes a \emph{universal ratio}, given by
\begin{equation}\label{universalmassratio}
\left( \frac{m_b}{m_f} \right)^2=\frac{\lambda_\ast}{3g^2_\ast}=R_m,
\end{equation}
as we approach the GNY QCP from the ordered side, which plays a crucial role in determining the scaling of the Drude peak within the critical regime [Eqs.~(\ref{Drude-general})-(\ref{mass-drude})].

\emph{Kubo formula}. We now compute the correction to
the OC at the GNY critical point separating a Dirac semimetal and a gapped ordered state at both finite frequency and temperature.
To this end, we use the Kubo formula relating the current-current correlation function to the conductivity, yielding the inter-band part
\begin{equation}
\sigma_{lm}^{\rm IB}(\omega)=2\pi\lim_{\delta\to0}\frac{\Im\Pi_{lm}(i\Omega\to\omega+i\delta,{\bf q}=0)}{\omega},
\end{equation}
and the Drude peak (at $\omega=0$ and any finite $T$)
\begin{equation}
\sigma_{lm}^{\rm D}(\omega)=-2\pi^2\delta(\omega)\lim_{\delta\to0} \Re\Pi_{lm}(i\Omega\to\omega+i\delta,{\bf q}=0).
\end{equation}
Here $\Pi_{lm}(i\Omega,{\bf q})$ is the Fourier transform of the current-current correlator in the space and imaginary time
$\Pi_{lm}(\tau,{\bf r})=\langle j_l(\tau,{\bf r}) j_m(0,0) \rangle$, while the fermionic current is
$j_l(\tau,{\bf r})=i\Psi^\dagger (\tau,{\bf r})\Gamma_0 \Gamma_l \Psi(\tau,{\bf r})$, and  $l,m$ are spatial indices.
For an isotropic system, the conductivity satisfies $\sigma_{lm}^{\rm IB, D}(\omega)=\sigma^{\rm IB, D}(\omega)\delta_{lm}$. Direct application of the Kubo formulae yields the OC of a two-dimensional noninteracting Dirac liquid, given by $\sigma^{\rm IB}_0(x)$ and $\sigma^{\rm D}_0 \left( x \right)$.

\emph{Interband optical conductivity}. We first consider the correction to the inter-band piece of the OC at the GNY critical point, which solely arises from the fermionic sector, since the critical bosonic excitations (composite objects of fermions) are charge neutral~\cite{comment}. We now use the fact that the bare (${\rm B}$) and renormalized (${\rm R}$) fermion fields are related through the wave-function renormalization ($Z_\Psi$) as $\Psi_{\rm B}=Z_\Psi^{1/2}\Psi_{\rm R}$, which in turn allows us to express the bare current-current correlator in terms of the renormalized one according to $\langle j_l j_m \rangle_{\rm B}=Z_\Psi^2\langle j_l j_m \rangle_{\rm R}$. Due to the gauge invariance the current does not receive any vertex renormalization. The leading order correction to the conductivity in the quantum-critical fan is then given by the wave-function renormalization $Z_\Psi$, computed in the quantum critical fan (i.e. for $g^2=g^2_\ast$) and at a finite temperature $T$. The wave function renormalization $Z_\Psi$, is ultimately related to the fermionic self-energy at zero external momentum, $\Sigma_f(i\Omega)$, explicitly evaluated in the Supplementary Materials (SM)~\cite{SM}, yielding
\begin{equation}\label{self-energy-finiteT}
Z_\Psi \left(x\right)=1-\frac{1}{2}g^2N_b\left[\frac{1}{\epsilon}+ b + f_1 \left(x\right)\right],
\end{equation}
after the analytical continuation $i\Omega\rightarrow \omega+i\delta$. The constant $b=[2-\gamma_E+\ln(4\pi)]/2 \approx 1.9769$, with $\gamma_E$ as the Euler-Mascheroni constant, and $f_1(x)$ is a purely real function of a real argument~\cite{SM}, satisfying $f_1(x\rightarrow\infty) \approx 2.18486$.
Now, using the form of the wave function renormalization, we find that the conductivity receives a nontrivial correction at the non-Gaussian GNY critical point
\begin{equation}~\label{interband-intermediate}
\sigma^{\rm IB}_\ast \left( x \right)= \left[ 1-N_b \; \frac{1+ \epsilon \; C(x)}{2N_f+4-N_b} \right]
\sigma_0^{\rm IB}\left( x \right),
\end{equation}
with $C(x)\equiv b + f_1 (x)$, which to the leading order in the large-$N_f$ expansion leads to the result quoted in Eq.~(\ref{interband-final}).

\emph{Correction to Drude peak}. Next we focus on the correction to the Drude peak in the vicinity of the GNY critical point. We compute this correction by approaching the QCP from the ordered (gapped) side of the transition, where both fermionic and bosonic excitations acquire a finite mass. However, they both tend to vanish with a universal ratio [see Eq.~(\ref{universalmassratio})] as the QCP is approached at $T=0$. The form of the Drude peak is then given by~\cite{SM}
\begin{equation}\label{Drude-general}
\sigma^D=8\pi^2N_f \delta(\omega)\int\frac{d^2{\bf k}}{(2\pi)^2}\frac{k_x^2}{E^2_{ k}}\left(-\frac{\partial n_f(E_k)}{\partial E_k}\right),
\end{equation}
where $n_f(z)\equiv[e^{z/k_BT}+1]^{-1}$ is the Fermi-Dirac distribution function, and $ E^2_{k}={k^2+m_f^2(T)}$ is the dispersion of the massive Dirac quasiparticles, ultimately yielding
\begin{equation}\label{Drude-interaction}
\sigma^D_\ast \left(x\right)=N_f F\left(\frac{m_f(T)}{k_BT}\right)\delta\left(x\right),
\end{equation}
where the universal function $F(y)$ reads
\begin{equation}
F(y)=\frac{\pi}{2}\int_0^\infty dk\,\frac{k^3}{(k^2+y^2)\cosh^2\left(\frac{1}{2}\sqrt{k^2+y^2}\right)}.
\end{equation}
The scaling of $F(y)$ is shown in Fig.~\ref{fig:funcF}.

To extract the correction to the Drude peak in the quantum critical fan, next we compute the \emph{thermal mass} of the fermionic field [$m_f(T)$] in this regime, where $T \lambda/m_j, T g^2/m_j \sim \sqrt{\epsilon} \ll 1$ for $j=f$ and $b$. An explicit calculation is shown in the SM~\cite{SM} and we find
\begin{equation}~\label{mass-drude}
\left(\frac{m_f (T)}{k_BT}\right)^2=\frac{\pi^2g^2_\ast  N_b}{\left[ R_m-1\right]}
=\frac{\pi^2 N_b}{6 N_f} \; \epsilon + {\mathcal O}\left( \frac{1}{N^2_f} \right),
\end{equation}
to the leading order in $\epsilon$ and $1/N_f$. Indeed, in the critical regime $m_f(T)/(k_BT) \sim \sqrt{\epsilon}$, since $g^2_*,\lambda_*\sim\epsilon$, see Eq.\ (\ref{FP:locationgeneral}). This result, together with Eqs.\ (\ref{Drude-interaction}) and (\ref{universalmassratio}), yields the interaction mediated correction to the Drude peak in the quantum critical fan, displayed in Eq.\ (\ref{Drude-final}), with the universal function $G(N_f,N_b)= \pi\sqrt{N_b/(6 N_f)}$ to the leading order in $1/N_f$~\cite{SM}. Most importantly, $F(y)$ is a monotonically decreasing function. Therefore, the weight of the Drude peak also decreases following a universal scaling function inside the quantum critical regime, while retaining its delta function shape. It is worthwhile noticing that $F(y)$ is not Taylor expandable close to $y=0$, since it originates from the derivative of the Fermi-Dirac distribution, which itself is not expandable about zero.

\emph{Long-range interaction}. Thus far we focused on a strongly interacting Dirac liquid, residing in the close proximity to a QCP, tuned by the short-range components (ones appearing in an extended Hubbard model) of the Coulomb interaction, which in a real system are always accompanied by the long-range tail. Although the long-range Coulomb interaction is believed not to alter the universal nature of this transition~\cite{juricic-semenoff, roy-dassarma}, it modifies the inter-band component of the OC in a $d$-dimensional Dirac system according to
\begin{equation}~\label{coulomb_general}
\sigma^{\rm IB}_{d}= \sigma^{\rm IB}_{0,d} \left[ 1+ C_d \; \alpha\left(r \right) \; \ln^{d-2} \left( r\right)\right],
\end{equation}
at $T=0$ and for $d=2$ and $3$. Here $\alpha=2 \pi e^2/v_F$ is the fine structure constant, a function of the running renormalization group scale $r=v_F \Lambda/(\hbar\omega)$, with $\Lambda$ as the ultraviolet momentum cutoff for Dirac fermions. $C_d$ is a universal number, with $C_2=(11-3 \pi)/6$~\cite{juricic-vafek-herbut} and $C_3=1/(3 \pi)$~\cite{roy-juricic-OC}. 
In two dimensions, the fine structure constant is \emph{marginally irrelevant}~\cite{vozmediano}, $\alpha(r) \approx 1/\ln(r)$ for $r \gg 1$, due to a logarthmically slow increase of the Fermi velocity ($v_F$) in the infrared, in agreement with experiment~\cite{geim-NatPhys2011}. Concomitantly the enhancement of the OC due to the long-range Coulomb interaction vanishes logarithmically slow as frequency $\omega \to 0$. Therefore, in a two-dimensional interacting Dirac liquid the correction to the OC at the GNY critical point arises \emph{solely} due to the strong coupling between the fermionic and bosonic excitations mediated by the finite-range interaction.

\emph{$(3+1)$-dimensions}. Finally, we briefly comment on the correction to the OC at the GNY critical point in three dimensional Dirac or Weyl systems, by focusing on the inter-band piece at $T=0$. In a three-dimensional non-interacting Dirac or Weyl liquid $\sigma^{\rm IB}_{0,3}= N_f e^2 \omega/(6 h v_F)$, while the fine-structure constant vanishes as $\alpha(r) \approx 3\pi/[N_f \ln(r)]$ for $r \gg 1$ and $N_f\gg1$~\cite{goswami, hosur, roy-juricic-dassarma}. The logarithmic correction to the OC in Eq.~(\ref{coulomb_general}) for $d=3$ stems from the violation of \emph{hyperscaling hypothesis} at the upper-critical dimension~\cite{roy-juricic-OC}. However, marginal irrelevance of the fine-structure constant conspires with the hyperscaling violation, leading to the following universal scaling of OC for $N_f \gg 1$
\begin{equation}~\label{coulomb3D}
\sigma^{\rm IB}_{\ast}= \sigma^{\rm IB}_{0,3} \left[ 1- \frac{N_b}{2 \epsilon} \; g^2_\ast + \frac{1}{N_f} \right].
\end{equation}
The part $\sim 1/N_f$ stems from the long-range tail of the Coulomb interaction. The quantum phase transition from Dirac or Weyl semimetal to an ordered phase in $d=3$ or $\epsilon=0$, driven by a short-range interaction, is \emph{mean-field} or \emph{Gaussian} in nature (since $g_\ast=\lambda_\ast=0$)~\cite{zinn-justin}; see also Eq.~(\ref{RG:zippedgeneral}). Thus, in $d=3$ the only correction to the OC that ultimately survives is due to the long-range Coulomb interaction, which is controlled via $1/N_f$, and the above expression with $g^2_\ast=0$ is an exact result to the leading order in $1/N_f$. This outcome is in stark contrast with the situation in $d=2$, where only the correction due to short-range Coulomb interaction survives  in an interacting non-Fermi liquid fixed point as $\omega\to 0$.

\emph{Discussion}. To summarize, we here present the quantum critical scaling of the OC at a relativistic fermionic QCP in two dimensions, as well as in the corresponding strongly coupled non-Fermi liquid. We show that both inter-band and Drude contributions decrease inside the critical regime in comparison to their counterparts in a non-interacting Dirac fluid, following universal scaling functions. This behavior can also be investigated numerically using quantum Monte Carlo simulations (see Refs.~\cite{wallin, sorensen, krempa, prokofiev, aurbach, katsnelson, Stauber}, for example), and possibly by using gauge-gravity or holographic dualities~\cite{zaanen-book}. In addition, the proposed $1/N_f$ scaling of OC [see Eqs.~(\ref{interband-final}),~(\ref{interband-intermediate})] can be tested numerically either (a) by externally changing the flavor number~\cite{hong-yao-2} or (b) by introducing inter-sublattice (hence without the infamous sign problem) third-neighbor hopping in a graphene-like model~\cite{bena}. Furthermore, our findings may be relevant in twisted bilayer graphene near so called `magic-angles' where sufficiently slow Dirac fermions can be susceptible toward interaction driven broken symmetry phases, since the Fermi velocity becomes $\sim 25$ times smaller than that in monolayer graphene, yielding nearly flat bands of Dirac fermions (with $N_f=4$)~\cite{cao-1,cao-2}, organic compound $\alpha$-(BEDT-TTF)$_2$I$_3$, residing at the brink of excitonic ordering~\cite{organic-DSM}, and given that the collisionless regime can be accessed in experiments~\cite{collisionless-exp}. Our analysis being restricted to the collisionless regime, cannot account for the \emph{smearing of the Drude peak} (since life-time of carriers $\tau \to \infty$). In future it will be interesting to investigate the quantum critical transport of a relativistic non-Fermi liquid in the collision dominated or hydrodynamic regime~\cite{RGJ-multicriticality, lars}, and find the crossover behavior of transport observables in a strongly interacting Dirac system. Finally, our findings may further motivate studies of the transport when critical fermionic and bosonic fluctuations are coupled in the vicinity of an extended Fermi surface, which can be germane for many strongly correlated materials such as cuprates, pnictides, and heavy-fermion compounds, for example~\cite{sachdev-book, berg-numerics-1, berg-numerics-2}.


\begin{thebibliography}{}
\bibitem{comment} This is strictly valid for Dirac semimetal-insulators or superfluid transitions. 
At a superconducting GNY QCP, the OC at GNY QCP receives additional contribution from gapless bosonic excitations of charge $2e$~\cite{SM}.
\end{thebibliography}
\end{document}